\documentclass{aa}
\usepackage{graphicx}

\newcommand{\msun}{M$_\odot$}
\newcommand{\lya}{Ly$\alpha$}

\newcommand{\bb}{$b_{435}$}
\newcommand{\vv}{$v_{606}$}
\newcommand{\ii}{$i'_{775}$}
\newcommand{\z} {$z'_{850}$}
\newcommand{\mic}{$\mu$m}
\begin{document}
\title{
Evidence for strong evolution of the cosmic star formation density at high redshifts.
\thanks{Based on observations collected at the European 
Southern Observatory, Chile, proposals 074.A-0233 and 076.A-0384}\\
}

\author{
    F. Mannucci,  \inst{1}
	H. Buttery,   \inst{2}
    R. Maiolino,  \inst{3}
    A. Marconi,   \inst{2}
\and L. Pozzetti  \inst{4}
}

\offprints{F. Mannucci \email{filippo@arcetri.astro.it}}

\institute{
     INAF, Istituto di Radioastronomia, Largo E. Fermi 5, 50125 Firenze, Italy
\and INAF, Osservatorio Astrofisico di Arcetri, Largo E. Fermi 5, 50125
     Firenze, Italy
\and INAF, Osservatorio Astronomico di Roma,  Via di Frascati 33,
     00040 Monte Porzio Catone, Italy
\and INAF, Osservatorio Astronomico di Bologna, 
     via Ranzani 1,  40127 Bologna, Italy 
}

\authorrunning{F. Mannucci et al.}
\titlerunning{Evolution of the SFD at high redshift}

\date{}

\abstract{
Deep HST/ACS and VLT/ISAAC data of the GOODS-South field were used 
to look for high-redshift galaxies in the rest-frame UV
wavelength range and to study the evolution of the 
cosmic star-formation density at z$\sim$7.
The GOODS-South area was surveyed
down to a limiting magnitude of about (J+Ks)$_{AB}$=25.5,
looking for drop-out objects in the \z\ ACS filter.
The large sampled area would allow for
the detection of galaxies that are 20 times less numerous and 
1-2 magnitudes brighter than similar
studies using HST/NICMOS near-IR data. 
Two objects had initially been selected as promising
candidates for galaxies at z$\sim$7, but were subsequently
dismissed and identified as Galactic brown dwarfs
through detailed analysis of their morphology and 
Spitzer colors, as well as through spectroscopic information. 
As a consequence, we
conclude that there are no galaxies at z$\sim$7 down to our limiting
magnitude in the field we investigated. Our non detection of
galaxies at z$\sim$7 provides 
clear evidence for a strong evolution of the luminosity function 
between z=6 and z=7, i.e. over a time interval of only $\sim$170~Myr. 
Our constraints also provide
evidence of a significant decline in the total star formation rate
at z=7, which must be less than 40\% of that
at z=3 and 40-80\% of that at z=6. 
We also derive an upper limit to the
ionizing flux at z=7, which is only marginally
consistent with what is required to completely ionize the Universe.


}

\maketitle

%---------------------------------------------------------------------------

\section{Introduction}
\label{sec:intro}

The sensitivity of new generation instruments, coupled with
8--10-m class ground-based telescopes and with the HST, allowed 
impressive progress in observational cosmology
during the past few years.

Several surveys aimed at searching
for high redshift galaxies have obtained large
samples of objects at increasing redshifts:
z$\sim$1 (e.g., Cowie et al. 1996), 
z$\sim$3 (Steidel et al. 1996, 2003), 
z$\sim$4 (Steidel et al. 1999, Hu et al. 1996), 
z$\sim$5 (Hu et al. 1999, Spinrad et al. 1998),
z$\sim$6 (Dawson et al. 2001; Fan et al. 2001; Stanway et al. 2003; 
Bouwens et al. 2004a, 2006; Dickinson et al. 2004; Mobasher et al. 2005),
up to the currently accepted record of 
z$\sim$6.6 (Hu et al. 2002;, Kodaira et al. 2003; Taniguchi et al.
2005).
Observations are approaching the interesting redshift range between z=6 and
z=10 where current cosmological models expect to find the
end of the reionization period and
the ``starting point'' of galaxy evolution (e.g., Stiavelli,
Fall \& Panagia 2004)

The results from all these surveys have had a tremendous impact on our
understanding of the cosmic history of star formation and on the 
reionization epoch (Gunn \& Peterson 1965).
As an example, the detection of high-redshift quasars by Fan et al. (2001)
allowed the measurement of the fraction of neutral gas 
at z$\sim$6 (Becker et al. 2001; Fan et al. 2002; 
Pentericci et al. 2002). The
discovery that this fraction is low but significantly non-zero 
(nearly the same as or larger than 1\%) 
puts severe constraints on reionization models. 
On the contrary, the 
detection of a high opacity due to ionized hydrogen in the WMAP data 
(Alvarez et al. 2006)
points toward a high reionization redshift (z$\sim$11). 
Thus, detecting objects at even higher redshifts could put strong
constraints on the properties of the
intergalactic medium at those redshifts (Loeb et al. 2005; 
Gnedin \& Prada 2004; Ricotti et al. 2004; Cen et al. 2005).

Most high-redshift galaxies from the previous surveys
were selected by the {\em dropout} technique, i.e., by
detecting the spectral break in the UV continuum blueward of the Ly$\alpha$\
due to intervening Ly$\alpha$-forest (see, for example, Steidel et al. 2003).
This technique is mainly sensitive to high-redshift
UV-bright galaxies, commonly named Lyman-Break Galaxies (LBGs). 
The use of different filters allows
the selection of different redshift ranges, from 
z$\sim$3 (U-dropout, Steidel et al. 1996, 2003), 
to z$\sim$6 ($i$-dropout, Stanway et al. 2003; Bouwens
et al. 2004a, 2006;  Shimasaku et al. 2005), 
z$\sim$7 ($z'$-dropout, Bouwens et al. 2004b;
Bouwens \& Illingworth 2006),
and even z$\sim$10 (J-dropout, Bouwens et al. 2005;
Mobasher et al. 2005);
see Hopkins (2004) and Hopkins \& Beacom (2006) for two recent reviews.

Up to a redshift of about 6, galaxies
can be detected by using optical data alone, as the Lyman break occurs at
$\lambda<0.85$\mic\ and the continuum above Ly$\alpha$ can be sampled, at
least, by the \z\ HST/ACS filter. 
This allowed for the detection of a large number
of objects at z=6, which provides evidence of strong evolution 
of the Luminosity Function (LF) of the LBGs between
z=6 and z=3. This is described as the brightening, with increasing cosmic age,
of the typical luminosity ($L^*$) of about 0.7 mag 
(Bouwens et al. 2006) or the increase in the comoving density ($\Phi^*$)
by a factor of 6 (Bunker et al. 2004).

At redshifts higher than 6 the use of near-IR images is mandatory,
which makes detection much more difficult. 
Deep J-band images from HST/NICMOS exist, but their
field-of-view is limited to a few sq.arcmin. 
Larger fields can be observed 
by ground-based telescopes, but at the expense of worse PSFs (about
0.5-1.0\arcsec\ FWHM instead of 0.4\arcsec\ for NICMOS/NIC3 and 
0.1\arcsec\ for HST/ACS)
and brighter detection limits.

Bouwens at al. (2004b) used one of these deep NICMOS fields in the 
Hubble UltraDeep Field (HUDF)
to detect $z'$-dropout objects. They detected 5 objects in a 5.7 sq.arcmin
field down to a limiting mag of H$_{AB}\sim$27.5, showing the possible
existence of a reduction in the Star Formation Density (SFD) at z$>$6. 
Recently, Bouwens \& Illingworth (2006) enlarged this search
to $\sim$19 sq.arcmin using new NICMOS data and improved data reduction. 
They detected  four possible z$\sim$7 objects, while 17 were expected
based on the z=6 LF. 
% Also, 4 out of the 5 source in the previous study turned
% out to be spurious objects. 
Also, they found that at least 2 of the 5 sources in their previous study 
were spurious.
These new results point toward the existence of a
strong reduction in the LF with increasing redshift at z$>$6.
In contrast, 
Mobasher et al. (2005), looking for J-dropout galaxies in the 
HUDF, detected one object that could be a massive post-starburst 
galaxy at z$\sim$6.5, formed at high redshift, z$>$9, (but see also Eyles et
al. 2006, and Dunlop et al. 2006, who suggest lower redshifts for this
object).
Deep near-infrared data were also used by Richard et al. (2006) to
examine two lensing clusters, and detect eight optical dropouts
showing Spectral Energy distributions (SEDs) compatible with 
high-redshift galaxies. After correcting
for lensing amplification, they derived an SFD
well in excess of the one at z=3, hinting at large amounts of star formation 
activity during the first Gyr of the universe. 

By extrapolating the LF observed at z$\sim$6 (Bouwens et al. 2006)
toward higher redshifts,
it is possible to see that those studies based on 
small-field, deep near-IR data are able to detect only galaxies 
that are much more numerous, and therefore fainter, than L$^*$ galaxies. 
Their faint
magnitudes imply that their redshifts cannot be spectroscopically confirmed
with the current generation of telescopes, and it is difficult to distinguish
these candidates from the Galactic brown dwarfs with the same colors.
As a consequence, these data cannot be used to confidently put strong
constraints on the amount of evolution of the LF at z$>6$: if all the
candidates are real star-forming galaxies at the proposed redshift, the 
SFD at z=7 and z=10 could be similar to that at z=6. 
If, on the contrary, none of them
are real, the upper limits of the SFD would imply a decrease of the SFD toward
high redshifts.
As a consequence, the evolution of the SFD at z$>6$ is 
not yet well constrained.
The knowledge of this evolution is fundamental for a good
understanding of the primeval universe, for example,
of the sources of reionization.

\bigskip

Here we present a study aimed at detecting
bright z=7 objects in a large (unlensed) field,
in order to measure the cosmic star-formation density at this redshift.
This is important for two main reasons: 
first, comparing these results for bright objects with those by Bouwens et al.
(2004b) and Bouwens and Illingworth (2006) for fainter $z'$-dropouts 
will allow us to study
different part of the LF. This is needed, for example, to
distinguish between luminosity and density evolution and  to
compare it with the dark matter distribution predicted by the cosmological
models. 
Second, the detections of relatively bright objects would allow
a more complete study
of the properties of these galaxies in terms of morphology, 
spectroscopy, and SEDs.
We will show that the absence of z$\sim$7 objects with
$M(1500)<-21.4$ in
the survey field means that the galaxy luminosity function evolved
significantly in the short time (about 170 Myr) between z=7 and z=6.
The epoch at z=7, about 750 Myr after the Big Bang, can therefore be
considered the beginning of the bulk of star formation.

Throughout this paper we use the concordance cosmology 
$(h_{100},\Omega_m,\Omega_{\Lambda})=(0.7,0.3,0.7)$ and AB magnitudes.

%----------------------------------------------------------------------
\begin{figure}     % figure 1
\includegraphics[width=9cm]{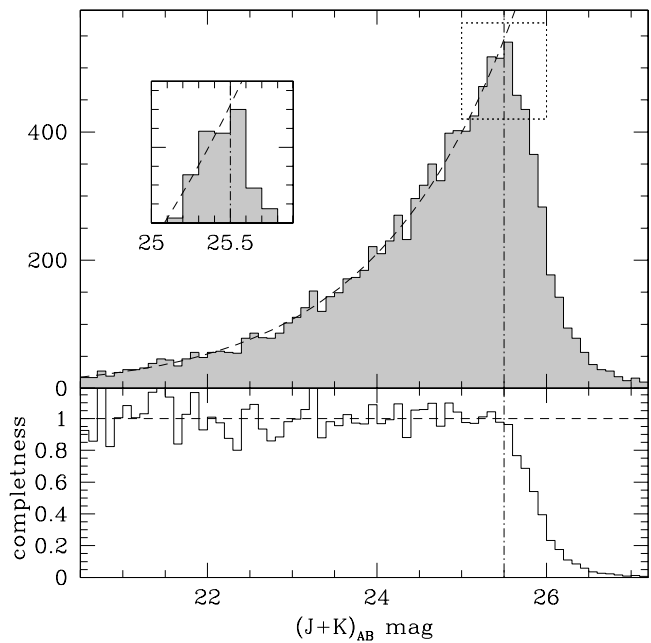}
\caption{
\label{fig:jklim}
{\em Upper:} histogram of the objects detected in the (J+Ks) image. 
The inserted box shows an enlargement of the dotted square. The dashed line
shows a polynomial fit to the histogram. The dot-dashed line shows the
used detection limit 
as derived from the statistics of the sky noise}. 
{\em Lower}: completeness of the
detections computed as the ratio between the fit and the histogram of the
upper panel.  The 6$\sigma$ limiting magnitude of (J+Ks)=25.5 
corresponds to a completeness of about 95\%. 
\end{figure}
%----------------------------------------------------------------------

\begin{figure}     % figure 2
\includegraphics[width=9cm]{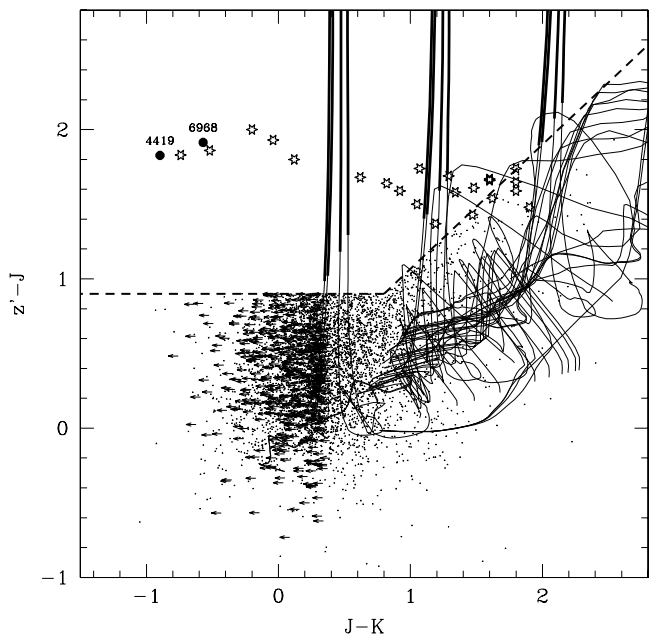}
\caption{
\label{fig:newsel}
Color-color selection diagram for z$>$7 galaxies. 
The solid lines show the variation in the colors with redshift
expected for local galaxies (Mannucci et al. 2001) 
and galaxy models by the Bruzual \& Charlot (2003).
Thin and thick lines show the expected colors for galaxies below and above
z=7, respectively.
The three groups of lines correspond to
three different amounts of extinctions (E(B-V)=0.0, 0.3, and 0.6, from left to
right), using the Cardelli et al. (1989) extinction law.
Stars show the expected positions of Galactic brown dwarfs
(ranging from  T8 to L1, from left to right). 
\newline
Small dots show the galaxies detected in the GOODS-South in all three
bands, while the arrows show the positions of the objects undetected in K. 
The dashed line shows the color threshold. Above this threshold only objects
with no counterparts either in \bb, \vv, and \ii\ or in the sum
\bb+\vv+\ii\ image are shown.
Large solid dots show the two objects discussed in Sect. 4.}
The labels show their entry number in our catalog.
\end{figure} 

%----------------------------------------------------------------------
\section{Observations and object catalog} 
\label{sec:obs}

The GOODS-South field, centered on the Chandra Deep Field South (CDFS),
is a region of about 10$\times$15 arcmin that is the subject of deep
observations by many telescopes. HST observed it with ACS using 
the broad-band filters \bb, \vv, \ii, and \z\ (Giavalisco et al. 2004a).
Most of the GOODS-South was observed in J, H, and Ks with the VLT/ISAAC 
(Labb\'e et al. 2003; Vandame et al. in preparation; 
see also Grazian et al. 2006). 
Many other deep observations are available for the GOODS-South field. 
The Spitzer satellite obtained
images of this field in all its bands, in particular in the IRAC bands 1 and 2,
corresponding to 3.6 and 4.5\mic. 

The main catalog is based on ESO/VLT J and Ks data
covering 141 sq.arcmin\footnote{To build the catalog we 
used the images of the public release version 1.0, 
while the photometry was computed on the images of the 1.5 version, 
available from October 2005}. 
The typical exposure time is 3.5h in J and 6h in K, with a seeing of about 
0.45\arcsec\ FWHM (see also Fig.~\ref{fig:morph}).
We used the J+Ks sum image to build the main object catalog: actively star
forming galaxies are expected to have a flat  
(in $f_{\nu}$ units) spectrum between J and K,
and therefore the use of the combined image is expected to improve
the object detectability.
The program Sextractor (Bertin \& Arnouts 1996) was used to extract the 
object catalog. We selected objects having flux above 1.2 times the RMS of the
sky over a 0.22~sq.arcsec area (10 contiguous pixels).
The part of the images near the edges were excluded 
as the noise increases because of telescope nodding. The final covered area
is 133 sq.arcmin.
The average 6$\sigma$ magnitude limit inside a 1\arcsec\ 
aperture is 25.5, estimated 
from the statistics of the sky noise. This agrees well with the
histogram of the detected sources as a function of the (J+Ks) magnitude
shown in Fig.~\ref{fig:jklim}, showing that this magnitude corresponds to
95\% of completness.
The main catalog comprises about 11.000 objects.

The colors of the objects were computed from the
photometry inside an aperture of 1\arcsec\ of diameter for the HST/ACS and
VLT/ISAAC images. 

%------------------------------------------------------------------
\begin{figure*}     % figure 3
\centerline{\includegraphics[width=18cm]{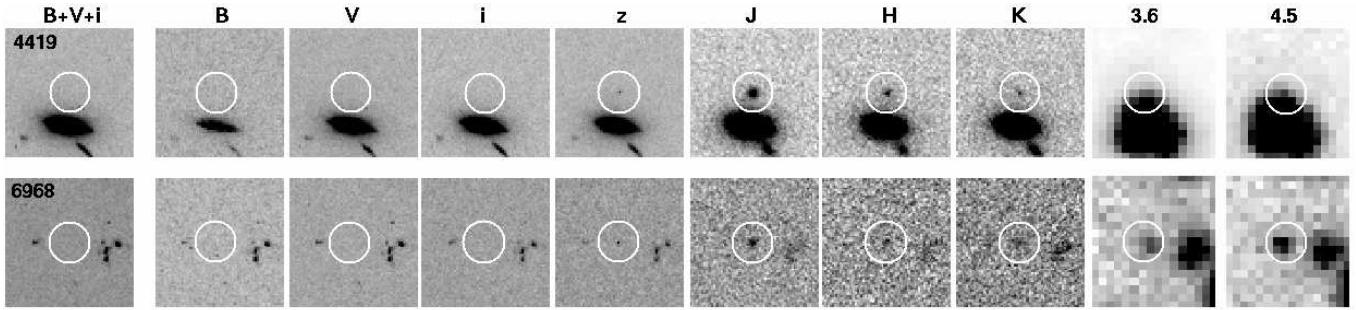}}
\caption{
\label{fig:images}
Images of the two candidates. Each box is 10 arcsec wide. The first panel
to the left shows the sum of the optical \bb, \vv\, and \ii\ ACS filters. 
The images in all the filters follow.
}
\end{figure*}
%-----------------------------------------------------------------

%-------------------------------------------------------------
\section{Candidate selection} 
\label{sec:newsel}

As discussed above, galaxies at z$\sim$7 can be selected as
z'-dropouts, i.e. objects with very red z'--J colors, indicating the
presence of a break, and with blue J--K colors, which disentangles 
high-redshift star-forming
galaxies from reddened foreground galaxies.
This is illustrated by the
z'--J versus J--K color diagram in Fig.2, where we plot the
colors expected for different classes of galaxies 
affected by various amounts of dust reddening
(from Bruzual \& Charlot 2003, and Mannucci et al. 2001) 
along with the
data from our sample. The figure shows that objects with
both red z'--J {\it and} red J--K color are likely to be
reddened galaxies at intermediate redshift. In Fig.2 the thick
tracks indicate the colors expected for star forming galaxies
at z$>$7, which are indeed separated from reddened galaxies on this
diagram.

Another class of possible interlopers are Galactic brown dwarfs.
Indeed, these stars can have blue J--K colors and
strong methane absorption shortward of $\sim 1\mu$m, strongly suppressing
the emission in the z' band and thus mimicking z'-dropouts of high redshift
galaxies. The coldest brown dwarfs are also undetected in all 
other optical bands.
This is illustrated in Fig.2, where stars show the colors of brown dwarfs
obtained by convolving the spectra in Testi et al. (2001) with the ACS 
and ISAAC filter transmission curves. 
These colors extend in the region of the diagram used for
selecting of z'-dropouts.
Summarizing, the use of near-IR colors alone is enough to exclude
low/intermediate galaxies, but not to distinguish true z$\sim$7 star-forming
galaxies from high-z QSOs or Galactic brown dwarfs. 

\bigskip

We selected objects that are undetected ($<1\sigma$) 
in \ii\ and in all bluer bands and that have $z'$--J$>$0.9 
and $\rm{J-Ks}<1.2*(z'-\rm{J})-0.28$, as shown in Figure~\ref{fig:newsel}. 
This region is larger than expected for galaxies at z$>$7, so
it is possible that a number of lower-redshift interlopers are present in the
catalog. This choice is motivated by not having detected any reliable candidate
(see below) and is therefore needed to exclude high-redshift galaxies
not being selected because of colors just outside an exceedingly
narrow selection area.

Despite the common name of ``dropouts'', we do not require that
our candidates are undetected in \z. 
In fact, z=7 sources can be quite luminous in this band for
two possible reasons: 
1) although the \z\ cut-on filter convolved with the ACS detector efficiency
has a response peaking at $\sim$8800\AA, its red wing
gathers some flux up to 1.05$\mu$m, and as a consequence, some flux at or
above the
Ly$\alpha$ line could be transmitted by the filter (indeed this is 
predicted by tracks
of galaxies at z$>$7 in Figure~\ref{fig:newsel}); 
2) if the reionization epoch is at z$>$7
(as the recent WMAP results and the spectrum of the galaxy at z=6.56
suggest, Bennett et al. 2003;  Hu et al. 2002), 
then the Ly-forest may still have some transmission, 
similar to what is observed in QSOs at z$\sim$6 (Pentericci et al. 2002).

\bigskip

An initial catalog was constructed by using the optical photometry
from the public GOODS catalog (Giavalisco et al. 2004a)
with the aperture photometry computed by Sextractor. This produced 
the selection of a few tens of $z'$-dropouts. 
These objects were checked by eye to 
remove spurious detections or objects whose photometry is seriously 
affected by bright nearby objects.
The photometry of the 6 objects that passed this check was measured again 
by using the IRAF/PHOT program in the same 1\arcsec\ aperture. 
Using a local sky computed in an annular 
region around the object, this program is expected to provide better 
photometry of faint sources. For 4 of these objects, the new $z'$--J
color is much bluer, below the detection threshold. This is an indication that
Sextractor overestimated the local sky near these sources. 

Two objects remain with colors that are
compatible with both star-forming galaxies at z$>$7
and Galactic brown dwarfs. 
Their images are shown in Figure \ref{fig:images}, while their
properties are listed in Table~\ref{tab:objs}
and discussed in the next section.

%------------------------------------------------------------------
\section{The two candidates}

The SED of the two candidates are 
compatible overall with being high-redshift star-forming galaxies.
The 9 available photometric bands, from \bb\ to Spitzer/IRAC 4.5$\mu$m,
can be used to derive photometric redshifts.
These two objects are contained in the GOODS-MUSIC 
catalog (Grazian et al. 2006), object 4409 with ID 7004 
and object 6968 with ID 11002 with 
photometric redshifts of 6.91 and 6.93, respectively.
Both objects have already been selected as $i$-dropout objects by other groups.
Object 4419 was selected by Stanway et al. (2003) (object SBM03\#5)
and further studied by Bunker et al. (2004) (ID number 2140), 
while object 6968 was selected by Eyles et al. (2006) (ID 33\_12465)
and by Bouwens \& Illingworth (2006).
Both objects were
identified as Galactic stars on the basis of the compact ACS morphology
and of the $z$JK colors. 
In the following
we use the morphologies, Spitzer data, and spectra to 
further investigate the nature of these sources.

Both candidates show very blue J--K colors and have $z'$--J$\sim$1.9. 
This blue J--K color could be due to 
starburst galaxies or faint AGNs at z$\sim$7 
with a UV spectral slope $\beta$ 
(with $f_{\lambda}\propto\lambda^{\beta}$  at $\lambda_{rest}=1500$\AA)
that is more negative than $-2$. 
These colors are also
typical of Galactic brown dwarf stars of
type T6-8, as can be seen in Fig. \ref{fig:newsel}.

In this section we present their properties in terms of 
the morphology in the optical and near-infrared images (Sect. \ref{sec:morph}), 
mid-infrared colors (Sect. \ref{sec:spitzer}),
and optical spectra (Sect.~\ref{sec:spectra}).

%------------------------------------------------------------------
% table 1
\begin{table}
\caption{High-redshift candidates}
\begin{tabular}{cccccc}
\hline
\hline
ID & \multicolumn{2}{c}{R.A. ~(J2000)~ DEC.} & \z & J & Ks\\
\hline
4419 & 03:32:38.8 & --27:49:53.7 & 25.10 & 23.27 & 24.17 \\
6968 & 03:32:25.1 & --27:46:35.7 & 26.39 & 24.48 & 25.05 \\
\hline
\end{tabular}
\label{tab:objs}
\end{table}

%----------------------------------------------------------------------
\begin{figure*}     % figure 4
\begin{center}
\includegraphics[width=8cm]{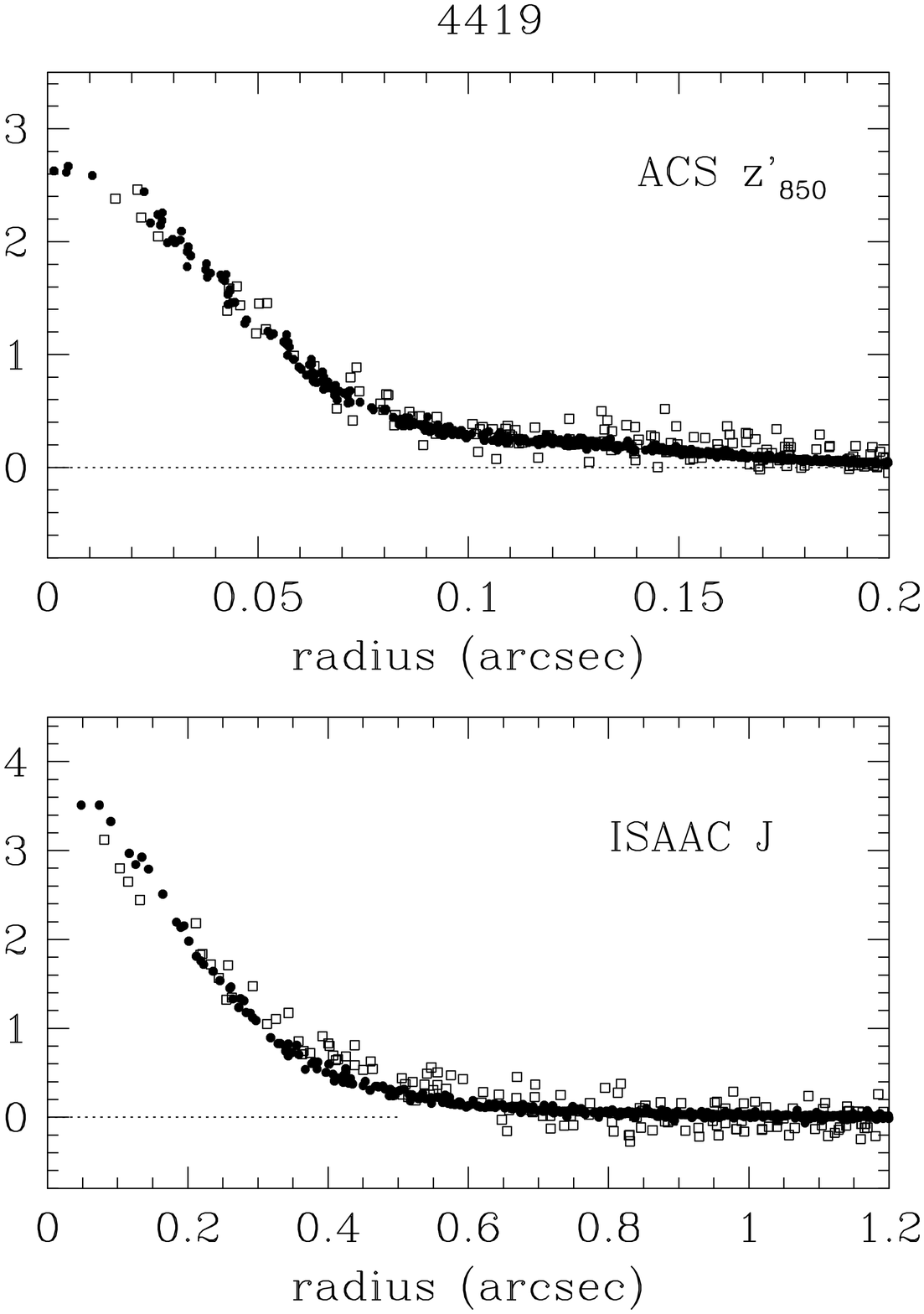}
\includegraphics[width=8cm]{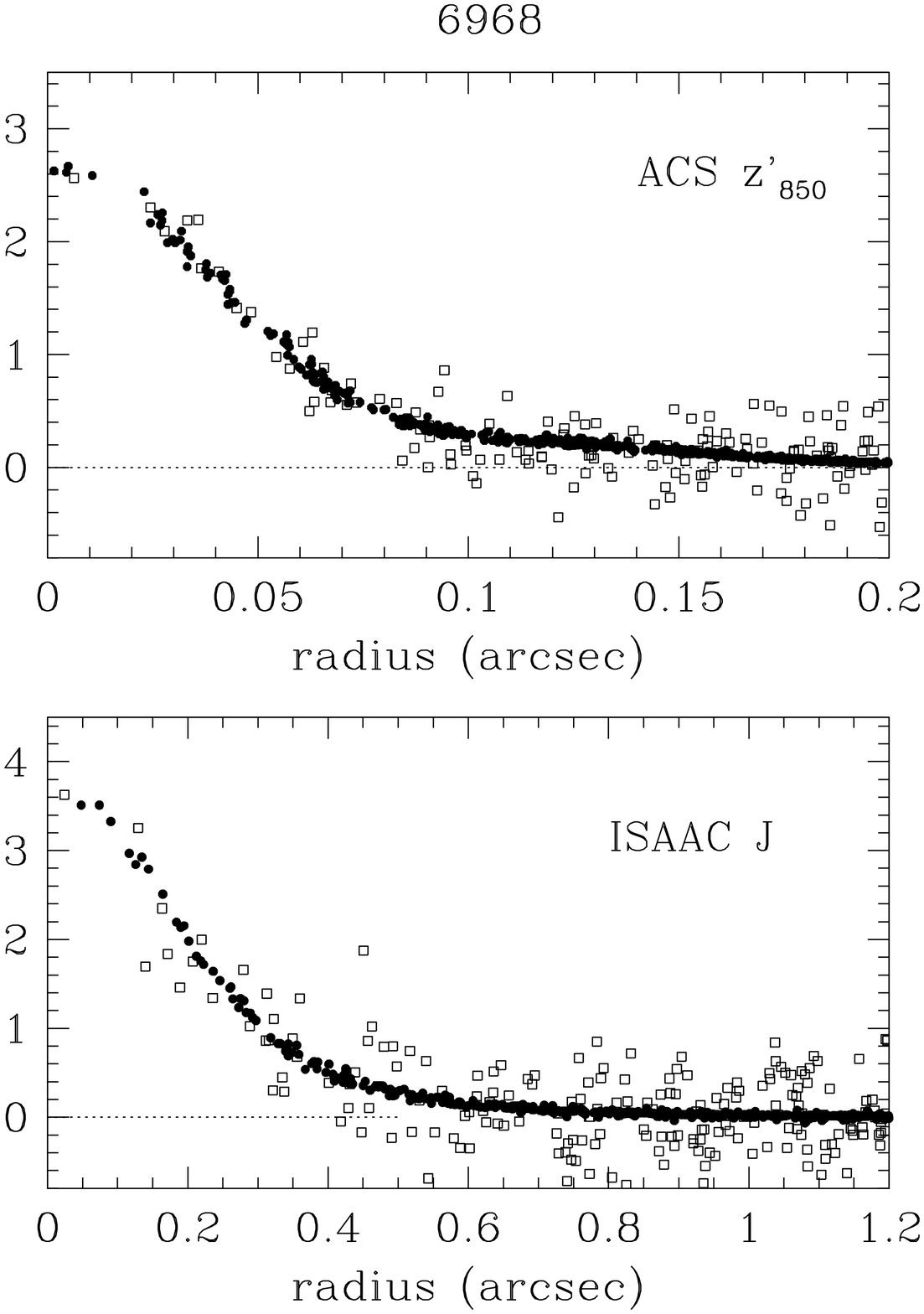}
\caption{
\label{fig:morph}
The observed luminosity profile of candidates 4419 (left) and 6968 (right)
are compared with the local PSF in the \z\ (upper panels) and 
J-band (lower panels) images. Black dots are the PSF derived from 4 nearby
stars, the white squares the luminosity profiles of the objects.
Both objects are unresolved.
}
\end{center}
\end{figure*}

%------------------------------------------------------------------
\subsection{Morphologies}
\label{sec:morph}

Both objects are detected in \z\ and J. We studied the morphology of these
objects in both bands: the HST/ACS \z\ images have a much higher
resolution (about 0.1\arcsec\ FWHM), 
but both objects are much fainter in this band. In contrast,
the VLT/ISAAC J band images have a lower resolution (about 0.45\arcsec),
but the higher signal-to-noise ratio per pixel allows a more accurate
study. In both cases we compared the object luminosity profile with the PSF 
derived from a few nearby point sources. As shown in Fig.
\ref{fig:morph}, both objects are consistent with being point sources, 
as no significant differences are seen with the local PSF.

%------------------------------------------------------------------
\subsection{Spitzer colors}
\label{sec:spitzer}

\begin{figure}     % figure 5
\includegraphics[width=9cm]{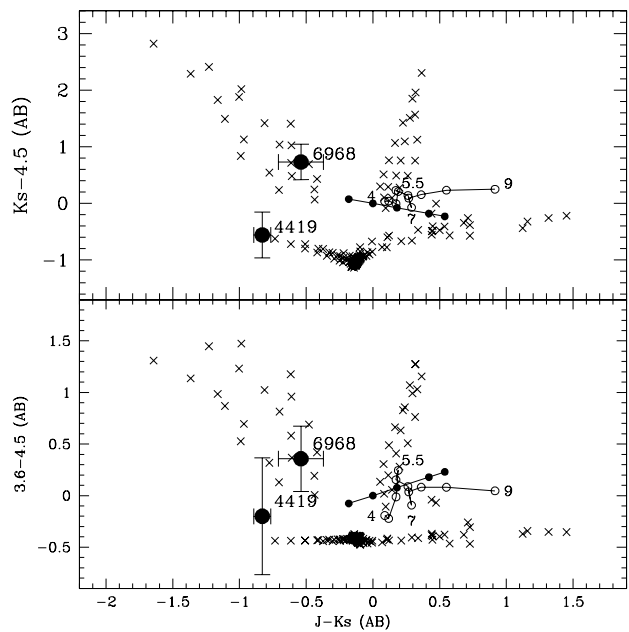}
\caption{
\label{fig:spitz2}
Near- and mid-infrared Spitzer/IRAC colors of the two candidates 
(black dots with error bars) compared with 
the expected colors of Galactic brown dwarfs (crosses, from Allard et al.
2001),
starburst galaxies at z$\sim$7 (connected black dots, representing galaxies  
with UV spectral slope
$\beta$=-2.3,-2.0,-1.7,-1.3, and -1.1, from left to right) 
and high redshift Seyfert1/QSOs (connected empty dots, for
redshifts between 4 and 9
with steps of 0.5, see the nearby redshift labels}).
The two candidates fall in the region occupied by the brown dwarfs.

\end{figure}

The near- to mid-IR colors can be used to investigate the nature of these
objects. 
Young starburst galaxies at $z\sim7$ are expected to be intrinsically quite
blue and show a flat spectrum above the Lyman limit. 
We computed the photometry of our candidates in the Spitzer/IRAC 
$3.6\mu m$ and $4.5\mu m$ images using 3\arcsec\ apertures,
in order to have aperture corrections 
similar to the those in the K band, i.e., about 0.42 mag (see, for
example, Eyles et al. 2005) 
and to compare the results with the near-IR data.
To compute the photometry of object 4419, the contribution 
by the nearby bright galaxy was estimated by convolving
the \z band image with the Spitzer PSF. This introduces a large additional
uncertainty.
Figure~\ref{fig:spitz2} shows the J--K vs. K--4.5$\mu$m and 3.6--4.5$\mu$m
color-color diagrams,
comparing the colors of both candidates with those of the Galactic stars
(derived from the brown dwarf models by Allard et al. 2001)
and of high-redshift galaxies and type 1 AGNs (Seyfert 1 and QSOs, 
Francis \& Koratkar, 1995).
It is evident that these diagrams can be used 
to distinguish compact galaxies from stars. 
Even if the Spitzer flux of 4419 is uncertain, because
it is affected by the nearby
bright galaxy, it appears that both objects show colors compatible with being
Galactic stars.
The flux in the Spitzer bands of object 6968 was already discussed by
Eyles et al. (2006), and they also concluded that they are compatible with
being a Galactic brown dwarf.

%------------------------------------------------------------------

%------------------------------------------------------------------
\subsection{Optical spectra}
\label{sec:spectra}

The two candidates were observed for 5.0 hr with the FORS2 spectrograph
on the ESO VLT in October 2005. 
The multi-slit facility allowed us to observe the two
candidates together with other objects having either very faint counterparts
in the \ii\ band or colors above the selection threshold when measured by 
Sextractor (see Sect.~\ref{sec:newsel}).  
We used
1\arcsec\ wide slits and the G600z grism, providing a dispersion
of 1.6\AA/pix and a resolution of about $\lambda/\Delta\lambda=1400$.
The covered wavelength range was between 0.75 and 1.05\mic.

In all cases, no significant emission line and spectral break 
were detected in this wavelength range. 
It should be noted that an optical spectrum of object 4419 have  already
been obtained by Stanway et al. (2004b) with DEIMOS on the Keck-II telescope,
but no spectral features were detected.

The emission-line detection limit was computed from 
the background noise. We obtain a 5$\sigma$ limiting magnitude 
for unresolved sources of
about $3\times10^{-18}$ erg~sec$^{-1}$~cm$^{-2}$ in the wavelength range
0.900-0.995\mic, corresponding to a redshift range for the \lya\ line of 
6.4$<$z$<$7.2. This limit holds for the part of the wavelength range
that is not covered by bright sky lines, which corresponds to about 88\%
of the total range.
Such a sensitivity would be more than adequate to detect the 
\lya\ emission from star forming galaxies at these redshift. This flux
can be estimated by assuming that the properties of the $i$-dropout galaxies
at z$\sim$6 also holds at these higher redshifts. The spectroscopic
observations of $i-$dropouts (Bunker et al. 2003; Stanway et al. 2004a;
Dickinson et al. 2004) have shown that color-selected galaxy have Ly$\alpha$
lines with rest-frame equivalent widths (EWs) of 20--30\AA, corresponding to 
$2-3\times10^{-17}$ erg~sec$^{-1}$~cm$^{-2}$ for the faintest of the two
candidates, well above the detection limit. It should be noted that
line-selected high-redshift galaxies show much larger EWs, on the order of
200\AA\ (e.g., Malhotra \& Rhoads 2002).
Similar values are derived for high redshift AGNs: scaling the \lya\ fluxes
in z$\sim$6 quasars observed by Fan et al. (2003) and Maiolino et al. (2004)
to the continuum luminosity of the current sample, we would obtain
fluxes on the order of $1-4\times10^{-17}$~erg~cm$^{-2}$~sec$^{-1}$.

\bigskip

As a conclusion, all of the four investigations presented above
confirm the identification of both
sources as Galactic stars. Also their luminosity is consistent with this
identification:
if they are brown dwarfs of type from T6 to T8, their absolute mag would 
be between M(J)=15.5 and M(J)=16.5 (Leggett et al. 2002), 
placing them between 200 and 700 pc from us. As a consequence, no z=7
object is present in the survey field above our detection limit.
  
%--------------------------------------------------------------
\section{Evolution of the luminosity function and of the cosmic 
star-formation density} 
\label{sec:lf}

The detection of no z$>$7 objects in the field can be used to place an upper
limit to the LF of these objects. 
To do this we need to
accurately estimate the volume sampled by our survey.
Two effects produce the dependence of this volume on the 
magnitude of the objects. First, fainter objects are detected in the
(J+Ks) image only at lower redshifts; second, 
only lower limits of the $z'$--J color can be
measured for objects selected in the (J+Ks) image and undetected 
in \z. This lower 
limit is above the selection threshold only for objects 
that are bright enough in the J band, 
while fainter objects can have a color limit below threshold.

\begin{figure}  % figure 6
\includegraphics[width=9cm]{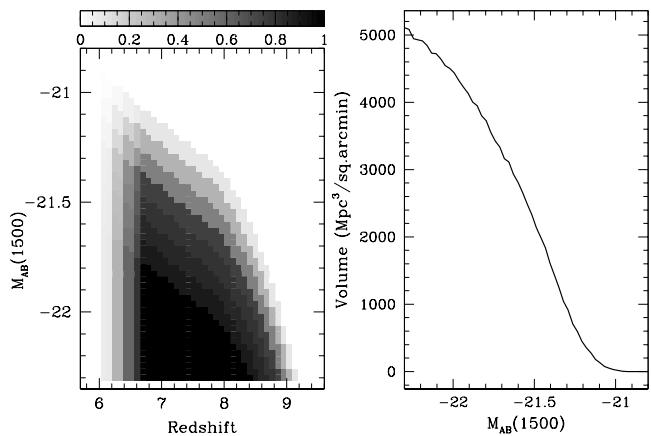}
\caption{
\label{fig:sampledvol}
{\em Left:} detection probability for objects of a given absolute magnitude 
and redshift derived by applying the selection in Sect.~\ref{sec:newsel}
and Fig.~\ref{fig:newsel}. The color code is described in the bar at the top. 
{\em Right}: the effective comoving volume $V_{eff}$ 
sampled at each absolute magnitude by the GOODS observations.
}
\end{figure}

The effective sampled volume $V_{eff}$ can be computed 
as a function of the absolute magnitude of the objects using

\begin{equation}
V_{eff}(M)=\int p(M,\rm{z})\frac{dV}{d\rm{z}}d\rm{z}
\label{eq:vol}
\end{equation}
\noindent
(Steidel et al. 1999; Stanway et al. 2003),
where $p(M,\rm{z})$ is the probability of detecting an object
of magnitude $M$ at redshift z using the selection in Fig.~\ref{fig:newsel},
and $dV/d\rm{z}$ is the comoving volume per unit solid angle at redshift z.
The results of this computation are shown in Fig.~\ref{fig:sampledvol}.

The detection probability $p(M,\rm{z})$ was obtained by computing 
the expected apparent magnitudes and colors of starburst galaxies
(modeled as $f_{\lambda}\propto\lambda^{\beta}$) 
of varying intrinsic
luminosity and spectral slope $\beta$, placed at different redshifts. 
Figure~\ref{fig:sampledvol} (left panel) shows the results of this computation.
The redshift sensitivity of our selection method starts at about z=6.7, 
followed by a peak at z=7 and a shallow decrease towards high redshifts. 
By integrating this function
over the redshift, we obtain the results
in the right panel of Fig.~\ref{fig:sampledvol}.

%----------------------------------------------------------------------
%----------------------------------------------------------------------

The limiting apparent magnitude ($\sim$25.5, see Sect.~\ref{sec:obs}) 
corresponds, at z=7, to
an absolute magnitude M(1500)=--21.4. Using the standard relation between 
UV luminosity and SFR (Madau et al. 1998), this value corresponds
to an SFR of $\sim$20\msun/yr. This value is adequate for sampling the brighter
part of the LF of LBG at any redshift. 
As an example, Stanway et al. (2003) found 9 $i$-dropout galaxies at z$\sim$6
with typical SFRs of 20--30 M\sun/yr, while the SFR of $L^*$ galaxies in 
Bouwens et al. (2006) is 9 \msun/yr. 
As a reference, the SFR corresponding to an L$^*$
galaxy in the $z\sim3$ sample by Steidel et al. (1996) is about 
15--20 M\sun/yr
(Giavalisco, 2002; Stanway et al. 2003).

\begin{figure}  % figure 7
\includegraphics[width=9cm]{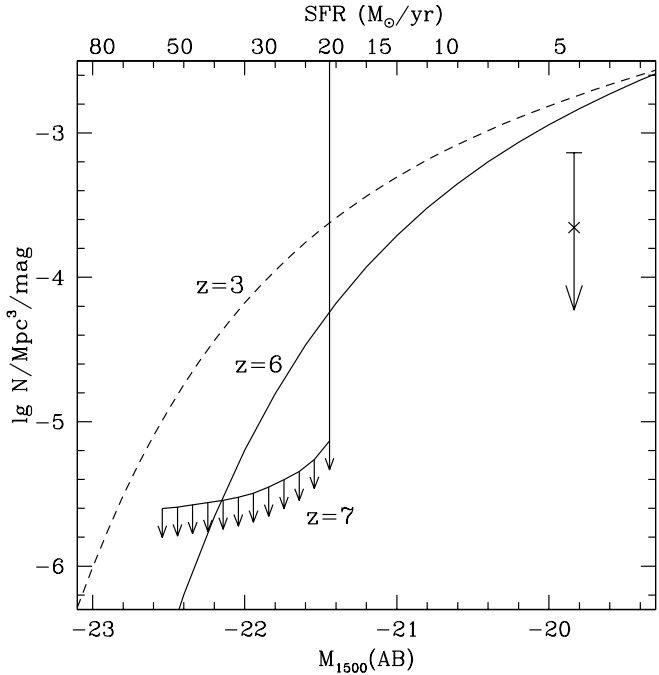}
\caption{
\label{fig:lf}
The limits to the density of z=7 galaxies (connected downward arrows)
are compared with the LFs of LBGs at z=3 (dashed line, Steidel et al. 2003) 
and z=6 (solid line, Bouwens et al. 2006, assuming M(1350)--M(1500)=0.10, as
for the typical z=3 LBGs). 
The upper limit with a cross corresponds to the object detected by
Bouwens et al. (2004b) in the HUDF and considered to be at z$\sim$7.
Data are plotted as a function 
of the absolute magnitude M(1500) at 1500\AA, while the upper axis shows
the corresponding value of the SFR. 
}
\end{figure}

This volume can be used to estimate an upper limit to the density of objects
at this redshift range. The limiting density for a given Confidence Level (CL) 
is given by
$$\rho(M,CL)=N(CL)/V_{eff}(M)$$
where 
$V_{eff}(M)$ is given by Eq. \ref{eq:vol} and $N(CL)$ is the 
maximum number of objects in the field corresponding to the limiting 
density at the chosen CL (2.3 objects for CL=90\%). 
Figure \ref{fig:lf} shows the 1$\sigma$ upper limits to the density
of objects, compared with the LF of the LBGs 
at z=6 from Bouwens et al. (2006) 
(a Schechter function with parameters
$M^*(1350)$=-20.25, $\alpha$=--1.73, and $\Phi_*$=0.00202~Mpc)
and at z=3 obtained by Steidel et al. (2003).
The LF at z=6 is still subject to large uncertainties because it is based on a
compilation of 500 $i$-dropouts coming from various fields with different
limiting magnitudes. As a consequence, cosmic variance could have different
effects in different parts of the LF. The faint-end slope is
particularly uncertain, as it is based on faint objects observed at low
signal-to-noise.  At this signal level, many interlopers
at lower redshift could be present in the sample, with the uncertainty on the
$i-z$ color causing galaxies to scatter into the selection region, hence
overestimating the number of faint sources.
It should also be noted that the VVDS collaboration 
(Le Fevre et al. 2005; Paltani et al. 2006)
studied the LF of the high-z galaxy population with 3$<$z$<$4 using a 
purely magnitude-selected spectroscopic sample, and found 
an unexpectedly large number of very bright galaxies, pointing toward the
possibility that the color selection could be effected by large incompleteness.

Besides these uncertainties, it is evident that our upper limits indicate 
an evolution of the LF from z=6, even if only 170 Myr have passed since then. 
In the no-evolution case, in fact,
we would expect to detect 5.5 objects, 
with a Poisson probability of no detection smaller than 0.5\%.
This is confirmed by the results in Bouwens et al. (2006) who
detected 7 objects at z=6 in 
the same luminosity range in the same field, 
in good agreement with our expectations.

Using a CL=90\% and assuming density 
evolution of the LF, we found that the normalization of the LF 
at z=7 must be at most 40\% of that at z=6.
The evolution of the LF from z=7 to z=6 could also be in luminosity rather 
than density, corresponding to a brightening of the objects with cosmic
time rather than to an increase in their number. 
This could be a better approximation of the real evolution
if the systems at higher redshifts tend to be smaller or less active than
the corresponding systems at lower redshift. In this case the minimum
evolution of the average luminosity compatible with our upper limits
at CL=90\% is about 0.22 mag. This evolution corresponds to a reduction 
of 20\% in the total SFD, obtained by integrating the LF 
assuming a constant faint-end slope $\alpha$.

Very similar results are obtained if the LF at z=6 by Bunker et al. (2004)
is used as a reference point. By analyzing a much smaller
sample of LBGs, these authors found that the shift in the LF
between z=3 and z=6 is consistent with density rather than luminosity
evolution. Using this determination, the ratio SFD(z=7)/SFD(z=6) 
is about 0.3 for the luminosity evolution and 0.8 for the density evolution.

Our limits can be compared with the LF at z=3 (Steidel et al. 1999).
Assuming luminosity evolution
we obtain a shift of $L^*$ about 0.9 mag, corresponding to 
SFD(z=7)/SFD(z=3)=0.42, a reduction of more than
a factor of 2.  
A pure density evolution would require SFD(z=7)/SFD(z=3)=0.05.

A strong reduction in the LF from z=6 to z=7 
is consistent with the tentative detection of one $z'$-dropout LBG at z=7
in the HUDF by Bouwens et al. (2004b) 
as revised by Bouwens \& Illingworth (2006).

%----------------------------------------------------------------------
\begin{table}	% table 2
\caption{Values of the UV luminosity density and SFD at z=7, where the 
limits are at 90\% confidence level
\label{tab:values}
}
\begin{tabular}{lcccc}
\hline
\hline
            &\multicolumn{2}{c}{$L>0.2L^*(z=3)$} &\multicolumn{2}{c}{$L>0.2L^*(z)$} \\
            & lg $\rho^a$  & lg SFD$^b$   & lg $\rho^a$   & lg SFD$^b$   \\
\hline
z=3~$^c$          &   ~26.17  &  ~--1.72   &  ~26.17   & ~--1.72 \\
z=6~$^d$          &   ~25.92  &  ~--1.98   &  ~26.10   & ~--1.80 \\
z=7, lumin. evol. & $<$25.76 & $<$--2.14 & $<$26.01 & $<$--1.89 \\
z=7, dens. evol.  & $<$25.52 & $<$--2.37 & $<$25.70 & $<$--2.20 \\
\hline
\hline
\end{tabular}
$(^a)$: erg/sec/Hz/Mpc$^3$ at 1500\AA\\
$(^b)$: \msun/yr/Mpc$^3$  \\ 
$(^c)$: Steidel et al. (1999)\\
$(^d)$: Bouwens et al. (2006)\\
\end{table}

%--------------------------------------------------------------------
\section{The evolution of the star formation activity}
\label{sec:sfd}

These findings on the relative evolution of the SFD can be compared with 
the results from other studies at lower and higher redshifts. 
The UV luminosity density $\rho_{1500}$ can be converted
to SFD by using the Madau et al. (1998) ratio.
Our upper limits to the density of galaxies at z=7 with (J+Ks)$<$25.5
can be directly
converted into an upper limit to the SFD contained in 
galaxies with M$_{1500}<$--21.44,  which turns out to be about 
$<$30\% of that at z=6 and $<$5\% of that at z=3. 
As a consequence, our data
show a strong reduction in SF activity in bright galaxies.

\begin{figure}  % figure 8
\includegraphics[width=9cm]{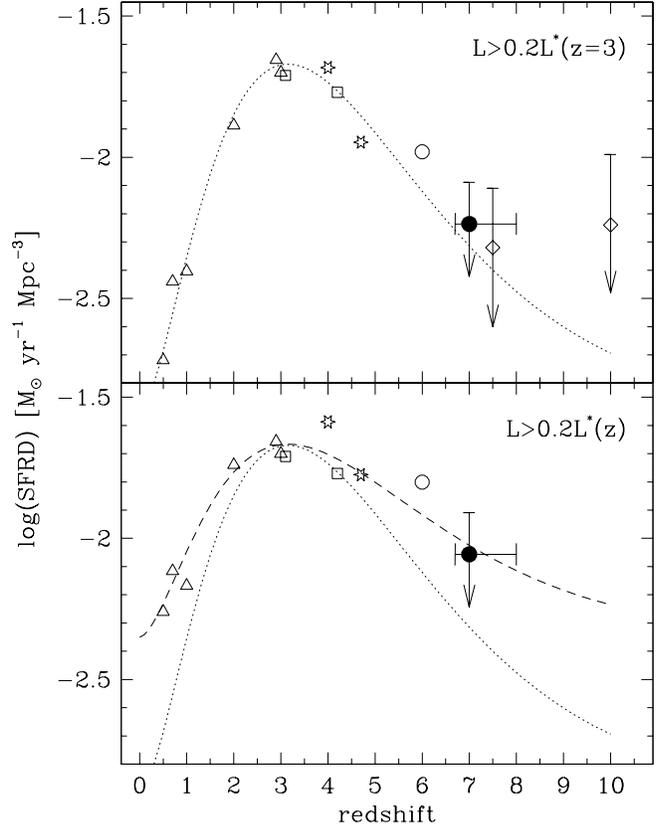}
\caption{
\label{fig:sfd}
Cosmic star formation as derived from 
rest-frame UV observations and with no correction for dust extinction.
The upper panel shows the results of integrating the LF for $L>0.2L^*(z=3)$,
the lower panel for $L>0.2L^*(z)$ (see text).
The black dot with error bars 
is obtained from the upper limit from the present work,
and the empty dots are obtained by integrating several published LFs
(triangles: Arnouts et al. 2005; squares: Steidel et al. 1999; stars: Ouchi
et al. 2004; circle: Bouwens et al. 2006) 
over the same interval.
The diamonds at z=7.5 and z=10 are from Bouwens et al. (2004b, 2005)
and are the only points not derived from an LF.
The original published values referred to $L>0.3L^*(z=3)$ and were corrected
to $L>0.2L^*(z=3)$ by assuming $\alpha=-1.73$ as observed at z=6.
The dotted lines shows an empirical fit to the data in the upper panel,
the dashed line to those in the lower panel.
}
\end{figure}

In Fig.~\ref{fig:sfd} we show the values of the SFD obtained 
by integrating the observed
LFs above a given luminosity threshold. The most interesting quantity, the
total SFD at each redshift, would be obtained by using a very low luminosity
threshold. 
Unfortunately, this would introduce large uncertainties due to the
unobserved part of the LF at low luminosities. For example, 
Steidel et al. (1999) observe their LBGs at z=3 down to $\sim$0.1$L^*$, while
for their value of the faint-end slope of the LF 
($\alpha=-1.6$), about half of the
total SFD takes place in galaxies below this limit. This implies that a
correction of about a factor of two is needed to obtain the total SFD from the
observed part of the LF. As for $\alpha<-2.0$,
the integral of the LF no longer converges, and this 
correction becomes even larger for more negative values of $\alpha$: it is 
2.4 for $\alpha=-1.73$ (as in Bouwens et al. 2006) and even 6 for
$\alpha=-1.9$, the most negative value of $\alpha$ compatible with the
i-dropouts in Bouwens et al. (2006). To avoid this additional uncertainty, it
is common to refer to the SFD derived from galaxies that where observed 
directly or by
using a small extrapolation of the LF. 
For the value at z=7, we plot the average between the two extremes of
pure luminosity and pure density evolution, 
while the upper value corresponds
to pure luminosity evolution.

The two panels of Fig.~\ref{fig:sfd} show the SFD as
obtained by integrating the observed LFs with two different lower limits. 
In the {\em upper panel} we show the results 
when considering, at any redshift,
the same range of absolute magnitudes ($M(1500)>$-19.32), i.e., 
limiting the luminosity to above 20\% of the $L^*$ magnitude
derived by Steidel et al. (1999) for their LBGs at z=3
($L>0.2L^*(z=3)$). 
The derived SFD is directly related to the total SFD in the case of 
pure density evolution, as the fraction of SFD from galaxies below 
the threshold would be constant. 
The use of the absolute limits of integration is 
very common (see, for example, Bouwens et al. 2006; Shiminovich et al. 2005). 
The resulting SFD appears to vary rapidly, increasing by more than a 
factor of 30 from z=0.3 to z=3 and then decreasing by a 
factor of 2.2 to z=6 and $\sim$4 to z=7.

In the case of luminosity evolution, the use of a constant 
limit of integration results in considering a variable fraction of LF.
For example, Arnouts et al. (2005) measured the UV LF at low redshift
and found a strong luminosity evolution, with $M^*$=--18.05 at z=0.055
and $M^*$=--20.11 at z=1.0. In this case, at low redshift
the limit of integration used above ($M(1500)>$--19.32) 
is even brighter than $L^*$ and only a small fraction of
the LF is integrated to obtain a value of the SFD. This is the
reason that the upper
panel of Fig.~\ref{fig:sfd} shows such a strong evolution at low redshift.
In the {\em lower panel} 
of Fig.~\ref{fig:sfd}, we use a variable
limit of integration, set to be $0.2L^*(z)$ at each redshift. This is more
suitable to reproducing the total cosmic SFD, as the luminosity evolution
appears to dominate both at low (Arnouts et al. 2005) and at high
redshifts (Bouwens et al. 2006). In this case, the obtained evolution is much
milder with an increase of a factor of 5 from z=0.3 to z=3 and a decrease
of $\sim$1.4 to z=6 and $\sim$2.5 to z=7. 
The resulting values of the UV luminosity density and 
of the SFD are listed in
Table~\ref{tab:values} for both limits of integration.

%--------------------------------------------------------------
%--------------------------------------------------------------

%-----------------------------------------------------------------------------
\section{Discussion}

The search for galaxies at even higher redshift, z$>$8, by looking for
J-dropout galaxies, faces another problem.
The absence of deep enough data at wavelengths 
longer than Ks make it impossible to select galaxies with the standard
color-color technique, and the selection relies on the J--Ks color alone.
As a consequence, this technique is prone to the presence of many
interlopers, and the constraints on the SFD at these redshifts 
are correspondingly weaker.
Bouwens et al. (2005) looked for z$\sim$10 galaxies by selecting 
J-dropouts in a deep HST/NICMOS field down to H$\sim$28. 
They detected 3 candidates, one of which is considered reliable.
It is currently not possible to investigate 
the nature of these 3 objects in greater detail.
This fact spoils the
possibility of using this result to put constraints on the SFD at z=10,
which is only constrained not to be larger than at z=6.
For this reason our results are broadly consistent with 
these limits.

On the contrary, our results are not consistent with the high value of the SFD
derived at z=6--10 by Richard et al. (2006). They measure a value 
of the SFD higher than at z=3, but they also warn 
against measuring SFD by using strongly clustered fields.

\smallskip

Even if the survey area is quite large for the 
obtained magnitude limits 
(25 times larger than the Hubble Deep Field, for example), 
the cosmic variance is 
still a potentially important concern. 
From the cross-correlation of the 
galaxies at this magnitude range it is possible to estimate 
(see, for example, Cresci et al. 2006) that the observed density can vary 
by about 20\% of the cosmic average because of this effect.  
Similar results were obtained by Somerville et al. (2004) on the basis of
cosmological simulations.
As a consequence, cosmic variance is not expected to be a dominant
effect.

\smallskip

All the data in Fig.~\ref{fig:sfd} are derived from UV observations
and, as a consequence, they are very sensitive to dust extinction,
as discussed by a large number of authors 
(see, for example, Meurer et al. 1999).
Variation in the dust content along the cosmic age is one of the effects
that could contribute to shaping the observed evolution of the SFD. 
The typical color of z=6 LBGs, represented by the UV spectral slope $\beta$,
is bluer at z=6 
($\beta=-1.8$ according to Bouwens et al. 2006, 
revised to $\beta=-2.0$ by Bouwens \& Illingworth 2006 
when using the new NICMOS zero points, 
and $\beta=-2.2$ for Stanway et al. 2005) than at z=3 
($\beta=-1.1$, Meurer et al. 1997, 
or $\beta=-1.5$, Adelberger \& Steidel 2000), pointing towards a
reduction in average extinction at high redshift of about a factor of two. 
As a consequence, the observed reduction in the SFD cannot be due to
an increase of the dust content. Actually, considering this effect 
would make the increase of SFD with cosmic time even more pronounced.

\smallskip

As we observe the bright part of the LF, we cannot exclude 
that the reduction in the number density of bright galaxies is 
not associated with an increase in that of the faint galaxies.
If, for
example, both $\Phi^*$ and $L^*$ vary together so that $\Phi^*L^*$ remains
constant, the resulting total SFD also remains constant. We cannot exclude
this effect, even if the upper limits to the galaxies with $L\sim L^*$
from Bouwens et al. (2004b) tend to exclude this possibility.

\section{Consequences on reionization of the primordial universe}

It is widely accepted that high redshift starburst galaxies can 
contribute substantially to the reionization of the universe.
Madau et al. (1999) (see also Bunker et al. 2004) estimated the amount of 
star formation needed to provide enough ionizing photons to the intergalactic
medium. By assuming an escape fraction of the photons $f_{esc}$ of 0.5 and a
clumping factor $C$ of 30 (Madau et al. 1999), 
we find that at z=7 the necessary SFD is

$$ \rm{SFD(needed)}\sim7.8\times10^{-2}~~~\rm{M}_\odot\rm{yr}^{-1}\rm{Mpc}^{-3} $$

\noindent
The observed total amount of SFD derived from UV observations can be derived 
by integrating the observed LF down to zero luminosity. 
Assuming luminosity evolution and integrating the LF down to $0.01L^*$,
we obtain 

$$\rm{SFD(observed)}=2.9\times10^{-2}~~~\rm{M}_\odot\rm{yr}^{-1}\rm{Mpc}^{-3}$$

\noindent
where about half of this comes in very faint systems, below 0.1$L^*$.
This value can be be increased up to 
$\sim5\times10^{-2}\rm{M}_\odot/\rm{yr}^{-1}\rm{Mpc}^{-3}$
assuming the 
steeper faint-end slope on the LF ($\alpha=-1.9$) compatible with the data
in Bouwens et al. (2006). 

The uncertainties involved in this computation 
(as the actual values of $f_{esc}$ and
$C$, the amount of evolution of the LF between z=6 and z=7, 
the faint-end slope $\alpha$ of the LF) 
are numerous and large for both SFD(needed) and SFD(observed).
Also, Stiavelli, Fall \& Panagia (2004) discuss how very 
metal-poor stars can overproduce ionizing photons and how a warmer IGM could
lower the required flux.
Nevertheless, the amount of ionizing
photons that can be inferred at z=7 from observations is less or, at most,
similar to the needed value. 
A measure of the SFD at even higher redshifts or tighter constraints
to the faint-end slope of the LF at z=6 can significantly reduce these
uncertainties and can 
easily reveal that it falls short of the required value.

\section{Conclusions}

The existing multi-wavelength deep data on the large 
GOODS-South field allowed us to search for z=7 star-forming galaxies
by selecting $z'$-dropouts.
The accurate study of the dropouts in terms of colors, morphology, and spectra
allows us to exclude the presence of any z=7 galaxy in the field above 
our detection threshold. We used this to derive evidence for the evolution 
of the
LF from z=7 to z=6 to z=3, and to determine an upper limit to the global 
star formation density at z=7.
These limits, together with the numerous works at lower redshifts,
point toward the existence of a sharp increase of the star 
formation density with cosmic time from z=7 to z=4, a flattening
between z=4 and z=1, and a decrease afterward.
The ionizing flux from starburst galaxies at z=7 could be too low to 
produce all the reionization.

The first star forming systems at z=7 appear to be too
faint to be studied in detail with the current generation of telescopes,
and the detailed comprehension of their properties will be
possible when the telescopes of the next generation come into use.

\begin{acknowledgements}
We are grateful to R. Bouwens 
for useful discussion and for having provided 
the catalog of his $i$-dropouts before publication, and to the referee, A.
Bunker, for very useful comments on the first version of the paper. 
We also thank the
Paranal staff for service observing and assistance during visitor
observations, and L. Testi for useful discussions about brown dwarfs.
\end{acknowledgements}

%\appendix

\end{document}